# Direct Experimental Evidence of Metal-Mediated Etching of Suspended Graphene

*Quentin M. Ramasse[1,*], Recep Zan[2,3], Ursel Bangert[3], Danil W. Boukhvalov[4], Young-Woo Son[4], and Konstantin S. Novoselov[2]*

[1]SuperSTEM Laboratory, STFC Daresbury Campus, Daresbury WA4 4AD, United Kingdom

[2]School of Physics and Astronomy, The University of Manchester, Manchester, M13 9PL, United Kingdom

[3]School of Materials, The University of Manchester, Manchester, M13 9PL, United Kingdom

[3]School of Materials, The University of Manchester, Manchester, M13 9PL, United Kingdom

[4]School of Computational Sciences, Korea Institute for Advanced Study, Seoul 130-722, Korea.

[*]To whom correspondence should be addressed, qmramasse@superstem.ac.uk

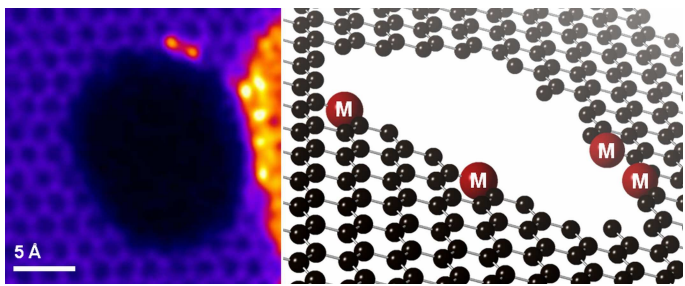

**Graphical TOC.** Atomic-resolution scanning transmission electron microscopy and *ab initio* modelling reveal how nano-holes are etched in single layer graphene sheets as a result of their interaction with deposited metal impurities.

**Abstract**

Atomic resolution high angle annular dark field imaging of suspended, single-layer graphene, onto which the metals Cr, Ti, Pd, Ni, Al and Au atoms had been deposited was carried out in an aberration corrected scanning transmission electron microscope. In combination with electron energy loss spectroscopy, employed to identify individual impurity atoms, it was shown that nano-scale holes were etched into graphene, initiated at sites where single atoms of all the metal species except for gold come into close contact with the graphene. The e-beam scanning process is instrumental in promoting metal atoms from clusters formed during the original metal deposition process onto the clean graphene surface, where they initiate the hole-forming process. Our observations are discussed in the light of calculations in the literature, predicting a much lowered vacancy formation in graphene when metal ad-atoms are present. The requirement and importance of oxygen atoms in this process, although not predicted by such previous calculations, is also discussed, following our observations of hole formation in pristine graphene in the presence of Si-impurity atoms, supported by new calculations which predict a dramatic decrease of the vacancy formation energy, when $SiO_x$ molecules are present.

*Keywords:* Graphene, Scanning transmission electron microscopy, EELS, dopants, single atoms, etching.

Since the first isolation of single layer graphite,[1] or graphene, a large body of research has been devoted to the remarkable electronic, structural and physical properties of this unique material, often with a view to utilising it for practical applications.[2] Almost by definition, however, the fabrication of any graphene-based device involves the incorporation of metal contacts, to exploit its thermal or electrical conductivity for instance. The use of either Au, Cr, Ti or Pd has been shown to dramatically affect the performance of the resulting devices and it follows that the choice of metal is therefore key to a successful design.[3, 4] Furthermore, metals like Zn,[5] Ni,[6, 7] Ag[8, 9] and Co[10, 11] or even non-metallic $SiO_x$[12] have all been used to tailor graphite and graphene into specific shapes such as nano-ribbons, by either oxidation or hydrogenation at elevated temperature, resulting in the formation of various by-products.[13] As a result, a number of recent studies have focused on the behaviour and interactions of deposited metal ad-atoms or nano-particles on graphene surfaces. Most of these are based on theoretical simulations such as Monte Carlo, molecular dynamics (MD) or density functional theory (DFT) to calculate the structure, bonding, and potential charge transfer of different adsorbed metal ad-atoms on materials.[14-18] Unfortunately, depending on the approach used for these calculations and in particular the choice of parameters and approximations, the results are not always consistent. Ni, for instance, was predicted to exhibit either strong[16, 19] or weak[18, 20] binding to graphene. As a general rule, however, transition metals (TM) have been predicted to bond covalently to the graphene surface resulting in significant lattice distortions whilst by contrast alkaline metals are ionically bonded to graphene and cause little distortion.[14, 21] In practice, it was recently demonstrated experimentally that instead of adhering to the free surface of graphene metals tend to cluster on top of the ubiquitous hydrocarbon-based contamination, which seems to indicate a very weak affinity between metals and graphene.[22] This weak TM-graphene interaction can be mediated by deliberately introducing vacancies (*i.e.* defective sites) into graphene sheets prior to deposition where some metal atoms can then be trapped[23-25] as a result of bonding rearrangements around the defects and a decrease of the adsorption energy barrier. In this case, it was recently calculated that with the exception of Au, whose interaction with graphene is consistently predicted to be weak, the presence of Al, Fe, Co and Ni metal atom impurities on or within the graphene layer can in turn lead to a dramatic reduction of the formation energies of further defects and therefore to the formation of large holes in the sheet. This would imply that graphene could be destroyed easily by the mere addition of metal atoms, a conclusion that has serious potential implications for graphene device fabrication.[26] It is therefore essential to devote renewed experimental attention to metal-graphene interactions in order to confirm or disprove this predicted deleterious behaviour and perhaps to propose new mechanisms whereby it could be alleviated. Transmission electron microscopy (TEM) is the ideal tool for such studies, especially with the mounting interest in so-called suspended devices, which eliminate substrate effects and thus exploit the intrinsic properties of free-standing graphene.[2] Indeed, the technique's ability to image and identify directly each and every atom in 2D materials has already played a significant role in the establishment of graphene as one of the most studied materials of the 21st century by providing arguably the most visually striking proofs of its existence.[27]

Here we present a systematic scanning transmission electron microscopy (STEM) study of the interaction of suspended single-layer graphene sheets with metals, namely Au, Cr, Ti, Pd, Ni and Al, as well as with Si (present in our samples as an unintentional impurity: see Materials and Methods), through a model system consisting of individual ad-atoms and/or aggregates of atoms (clusters) deposited on free-standing graphene. The effects on the structure of graphene were observed at ambient temperature in ultra high vacuum ($10^{-9}$ torr) and at a primary beam energy (60 kV) well below the knock-on damage threshold for carbon,[28] *i.e.* in conditions that are expected to allow for safe, prolonged, observation of the material without altering its structural integrity. A combination of chemically-sensitive Z-contrast imaging and electron energy loss spectroscopy (EELS) allowed us to confirm that in spite of these specific environmental conditions, etching does take place in the presence of all deposited elements with the exception of Au. Etching systematically initiated at the edges between clean graphene areas and the macro-molecular contamination layers where the metal clusters and impurity atoms tend to sit. Based on further theoretical results, we suggest that this behaviour is due to the local oxidation of the metallic ad-atoms, the oxygen activation energy barrier being possibly overcome by local heating as a result of energy transfer from the beam followed by C-atom dissociation through C-O formation.

Although only on a model system, the observation and elucidation of this metal-mediated etching behaviour is an essential result at a time when graphene is moving from the laboratory to the factory floor. A better understanding of the properties of metal contacts on suspended layer graphene is essential for device fabrication, patterning and improving performance and our results point to the need for more systematic studies of nucleation and coverage in order to determine optimal contacts.[3, 4]

**Results and Discussions**

The monolayer graphene sheets used for this study were grown by chemical vapour deposition (CVD) on a copper substrate according to the method described by Li *et al.*[29] and metals were introduced onto the graphene membranes by means of evaporation, prior to STEM measurements. Figure 1 shows low magnification HAADF images of Au (a), Cr (b), Ti (c), Pd (d), Ni (e) and Al (f) deposited on single layer graphene sheets (as verified using electron diffraction in a conventional TEM prior to deposition: see supplementary material, figure S1). The same amount of metal (2 Å) was evaporated in each case for accurate comparison. Consistent with recent observations, the deposited metals have formed clusters located exclusively on top of the ubiquitous hydrocarbon contamination (residues from the transfer, or contamination due to exposure to air) partially covering the surface of graphene.[22] No such clusters were observed directly on clean single-layer graphene (the darker patches on fig. 1) throughout this study, irrespective of the preparation method. A similar amount of Au was deposited on exfoliated single layer graphene (not shown here) resulting in a similar behaviour to the CVD-grown samples, on which we will concentrate for the rest of this study. This lack of adherence of transition metal atoms on pure graphene, or in other words their apparent high mobility on clean graphene surfaces, illustrates how significantly stronger the metal-metal interactions are than the metal–graphene interactions.[22, 30] Indeed this behaviour was predicted by DFT calculations[31] and MD simulations,[32] which suggested that clustering is more energetically favourable for transition metal atoms than remaining isolated. By contrast alkali metals are expected to form 2-dimensional continuous films on top of the graphene surface.[32]

In spite of identical deposition conditions the physiognomy of the samples varies significantly. Au clusters (fig. 1a) are larger and denser and as a result coverage is sparser than for the other metals. Pd (fig. 1d) and Ni (fig. 1e) also agglomerate into well-defined metallic nano-particles, albeit of much smaller sizes than Au, resulting in a more uniform coverage of the sample. Cr (fig. 1b) and particularly Ti (fig. 1c) and Al (fig. 1f) exhibit a much higher fractional coverage with loose, flat, atomic aggregates.

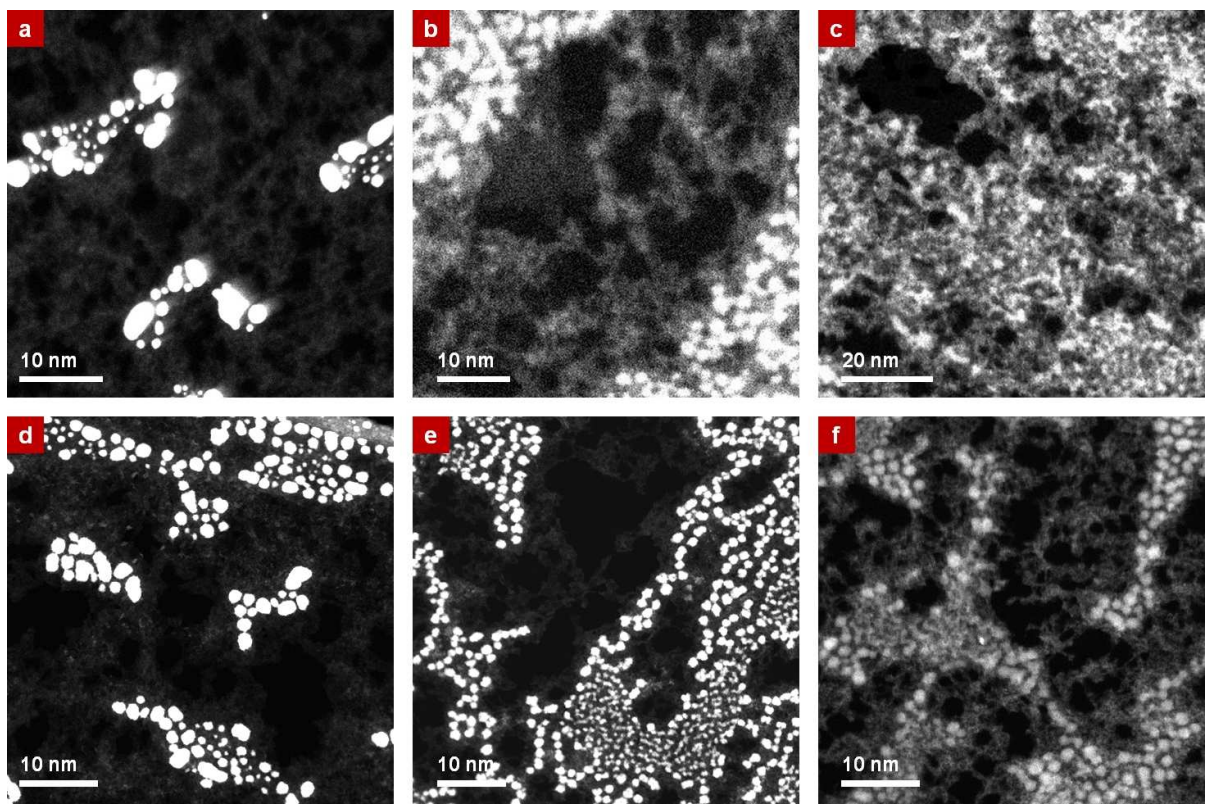

**Figure 1.** Low magnification HAADF images of metals on monolayer graphene show an overview of the metal distribution for: (a) Au; (b) Cr; (c) Ti; (d) Pd; (e) Ni and (f) Al. The contrast and intensity were adjusted to reveal areas of clean graphene (dark patches) and 'contamination' layers (light grey patches) where the metals (white clusters) sit preferentially.

The propensity of the latter to oxidise into alumina may explain the aspect of that particular sample as the deposited Al may have oxidised during sample transport from the deposition chamber to the microscope or upon contact with the hydrocarbon layer. A similar argument can be made for Ti and Cr, which are commonly used as a precursor for the fabrication of Au contacts on graphene: the better coverage of the sample after deposition of those two metals observed here may be an illustration of their effectiveness for such applications. Although none were applied here, surface pre-treatments such as hydrogenation (or oxidation) have been shown to affect the adherence of metals to the samples, resulting for instance in smaller Au clusters and a more uniform coverage which in turn is easily degraded by beam-induced coalescence of the clusters under the electron beam.[30]

Although as can be seen in the overview images of fig. 1 the metal clusters sit preferentially in the middle of the hydrocarbon contamination, after a few scans of the electron beam at mid- to high-resolution (for high signal-to-noise images each scan can take up to 30s), some of the clusters and/or individual atoms can be dragged by the beam to the edge of the contamination layer. Figure 2a shows an HAADF image of such a Ni cluster positioned at the very edge of a contamination area. Individual Ni atoms can be also seen to form a raft above a region of clean single layer graphene at the edge of the particle. While some carbon chains may still be present immediately under these Ni atoms, such a configuration offers a great proximity between pure graphene and the metal atoms. After a few additional scans of the beam, a hole has formed (fig. 2b), decorated by individual Ni atoms as evidenced by the clear Ni $M_{2,3}$ signature in the EEL spectrum (fig. 2c), acquired by placing the electron probe exactly on top of the bright atom marked on fig. 2b. This hole formation is obviously dynamic and additional stills from a time series of over 90 consecutive images illustrate the process further (see supplementary material, fig. S2). After the initial hole formation, individual Ni atoms are observed to jump onto the exposed edge before the hole is further enlarged, producing bright horizontal streaks in the images (figure 2a) as they are being captured at different positions by the beam as it is being rastered in a line. They can only be imaged once in a more stable position at the edge of the hole, such as on figure 2b. A strong indication that the drilling process is indeed metal-mediated arises from the observation that when no Ni atom is decorating the hole, the latter merely re-shapes dynamically (as expected from earlier reports[33]) but does not grow further in size (see supplementary material, fig. S3). In other words, the drilling stabilises when the local reservoir of metal impurity atoms is exhausted and until more Ni atoms are drawn towards the energetically unstable edges of the hole *via* surface diffusion thanks to the high mobility of single metal impurities on single layer graphene.

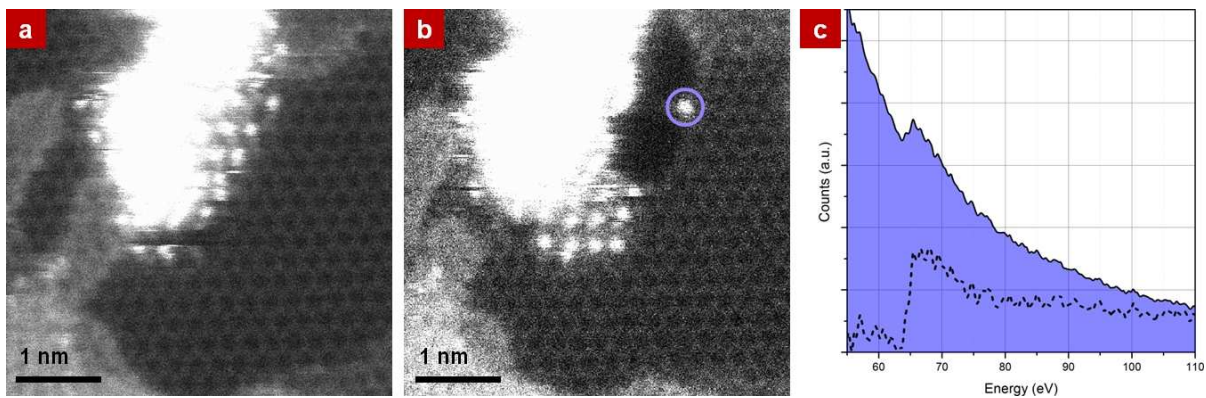

**Figure 2.** (a) HAADF image of a Ni cluster sitting on the very edge of the hydrocarbon contamination layer. Single Ni atoms have been dragged by the beam from the cluster and are in contact with the graphene monolayer. (b) After a few more scans, a hole has formed, whose edges are decorated by single Ni atoms, identified by their Ni $M_{2,3}$ EELS signal. (c) EELS spectrum acquired by positioning the beam for 1s on the bright atom circled in (b).

This behaviour (migration of the metal atoms under the beam to the edge of the contamination layer, drilling and hole enlargement) was reproduced identically when imaging single layer graphene samples on which palladium (fig. 3a), titanium (fig. 3b) and aluminium (fig. 3c) had been deposited, while in an earlier report a similar process had probably been at play but not recognised for the interaction of Cr with graphene.[22] In each case, the nature of the atoms decorating the edges of the newly formed holes was confirmed by placing the electron probe directly on top of them and recording an EELS spectrum, as shown on fig. 3.

Metals have been used as catalysts for patterning of graphene devices in hydrogen or oxygen flow at high temperatures,[6, 9, 11] and the addition of Ni in particular was predicted to lead to very low defect formation energies when interacting with single layer graphene.[26] However, neither gas environment nor high temperatures were used in our study, which is to our knowledge the first experimental evidence of electron beam-induced drilling of graphene through its direct interaction with metals. D. W. Boukhvalov and M. I. Katsnelson predicted this destructive behaviour (specifically for Fe, Co, Ni and Al) by calculating a drastic lowering of the formation energy for mono- and di-vacancies in single layer graphene when metal ad-atoms are present on the graphene surface.[26] The precise mechanism they propose in their *ab initio* calculations assumes a direct contact between a metal atom and the surface of free-standing graphene before the formation of the defects: as noticed above, some direct contact can be observed at the very edge of the contamination layer after a few scans of the electron beam, as illustrated on fig. 2a. While our observations seem to provide a direct experimental evidence of this metal-mediated destruction of graphene, it is important to consider other possible reasons for the severe drilling behaviour we observed.

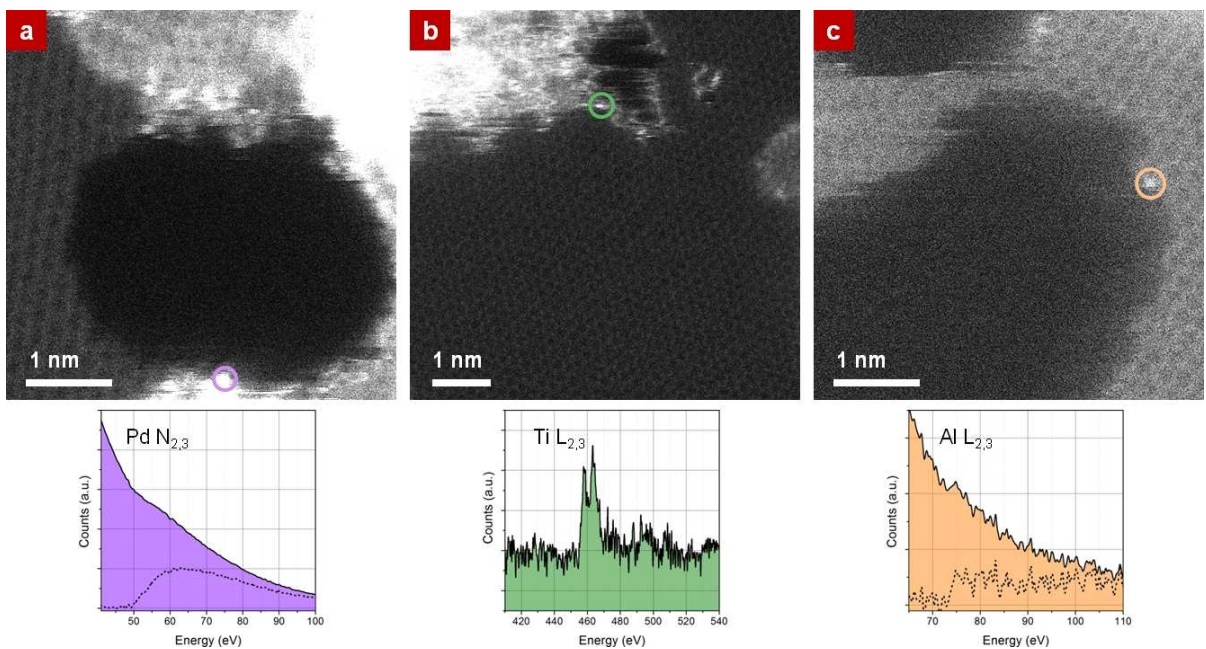

**Figure 3.** HAADF images of holes formed in monolayer graphene through metal mediated etching for (a) Pd, (b) Ti and (c) Al. EELS spectra acquired by positioning the beam for 1s on the atoms circled on the images are shown below. The dotted lines correspond to the EELS signal after background subtraction with a power law.

Although all are transition metals, these elements have been predicted to interact with graphene quite differently. Palladium has been used as an electrical contact in device fabrication for many years[34] and most recently in graphene devices[35] because of its lower cost compared to gold. Although it has a full d-shell, it is expected to bond covalently to graphene with a reasonably high adsorption energy.[14, 19] Recent DFT calculations also suggest that Pd atoms have a strong tendency to form three-dimensional rather than planar clusters on graphene, which indicate a weak Pd-graphene interaction[36] as evidenced by our observation, see fig 1d. Of all the metals used for this study, Ti is predicted to have the strongest interaction with graphene, bonding to its surface *via* chemisorption.[17, 19] Finally, Al is predicted to have ionic bonding to the graphene surface, similarly to the case of I-III metals, which unlike transition metals[14] are not seeing a significant modification of their electronic state. Furthermore, Al-doped graphene has been shown to be a promising material for hydrogen storage[37] while the deposition of a layer of aluminium oxide can be used as a gate insulator in graphene device fabrication.[38]

A constant trait of all those elements is their propensity to form oxides, which suggests that oxidation could be playing a major role in the effects we are observing. This theory could be further strengthened by the fact that by contrast no hole-forming was observed on Au-deposited samples (see supplementary material, fig. S4 and, for instance, Zan *et al*.[30]), Au being of course not prone to oxidation except in very specific circumstances.[39] Nevertheless, even in the mono-vacancy formation model for the destruction of graphene proposed by Boukhvalov *et al*.,[26] Au is not expected to lead to a major loss in stability as its defect formation energy remains high so this observation cannot definitively point to a role of O in this etching process. As the depositions were conducted in thermal and e-beam evaporators where the vacuum is at least $10^{-7}$ torr, all metals being degassed prior to evaporation, the most likely sources of oxygen, should any be involved in this process, are therefore: the oxidation of the metal clusters and/or the retention of O by the hydrocarbon contamination during transport of the samples from the deposition chamber to the microscope; or a relatively high partial pressure of O in the microscope column. All samples were systematically left within the microscope vacuum for several days to ensure perfect thermal and mechanical equilibrium during observation: after such long waiting periods the slight pressure increase in the column due to sample insertion had subsided and the sample chamber was systematically at its base pressure of $<5\times10^{-9}$ torr. Sample degassing can be thus considered as an unlikely source of O. Graphene etching in oxygen environment is well-documented and the energy required for the oxidation of graphene is expected to be low.[40] Therefore, should a high partial pressure of O in the chamber be responsible for the observed hole formation it should be occurring everywhere, not only at the edges of the hydrocarbon contamination. Furthermore, at 60kV and in otherwise identical conditions perfectly clean patches of graphene were imaged by scanning the beam repeatedly for over an hour without any drilling. The hole formation mechanism we report here must therefore be either solely metal-mediated as in the model from Boukhvalov *et al.*,[26] or involve metal adatoms and oxygen from either oxidised metal clusters or oxidised surface contamination.

As a control experiment we studied a 'pristine' graphene sample, *i.e.* a single layer graphene sheet produced and processed in identical conditions but on which no metal was deposited. As mentioned previously, in the conditions used for our observations pristine single layer graphene patches can be imaged without any visible damage for extremely long periods of time, and with very large electron doses. Some carbon surface contamination may occasionally diffuse into the field of view depending on the area being observed (the C support film may act as a 'contamination reservoir'), but drilling or etching was never noticed when imaging clean areas of our samples. Although every attempt was made at obtaining extremely clean single layer graphene samples prior to metal deposition, Si contaminants, either in the form of relatively large $SiO_x$ clusters or of small single Si atoms are common[41] and were readily observed in our samples.

Substitutional Si impurities were found to be very stable. Figure 4a shows an HAADF image of a such a Si impurity, identified by EELS by placing the beam directly on top of it (figure 4b). Several such datasets were acquired sequentially, with the Si impurity atom always appearing on HAADF images recorded immediately after the 2s EELS acquisition took place, proof of the great stability of this structure. However, when imaging continuously the edge of a hydrocarbon contamination layer containing some Si impurities (identified by EELS), we were able to observe even on such 'pristine' samples the formation of a hole in the graphene sheet, although not as readily as with metal-deposited samples.

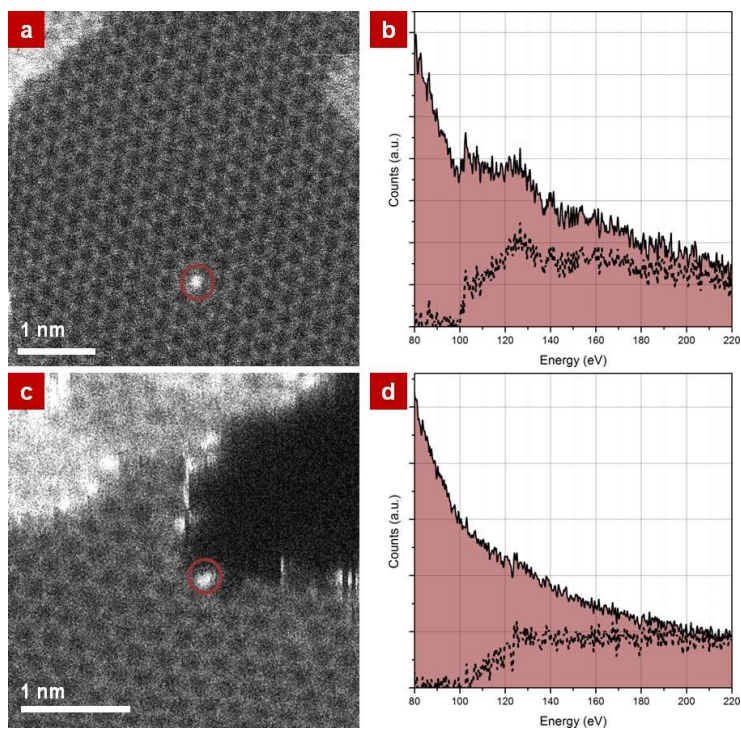

**Figure 4.** (a) HAADF image of a single substitutional Si atom within a graphene monolayer. An EELS point spectrum (b) acquired by positioning the beam for 1s on the atom confirms it is Si. This defect is extremely stable as it was possible to record several successive such datasets, the atoms remaining in position throughout. (c) HAADF image and corresponding EELS point spectrum (d) of a Si atom decorating a hole, just formed at the edge of the hydrocarbon contamination layer. The signal after background subtraction with a power law model is shown as the dotted line.

As in the metal case, the edges of the thus-formed hole were subsequently decorated by single atoms (figure 4c), identified as Si atoms by EELS (figure 4d). Again, the process was observed to continue until the local reservoir of Si impurity atoms is exhausted, at which point the drilling is halted and only a dynamic reshaping of the edges of the hole can be observed under the beam.

$SiO_x$ has been shown to have potential applications in tailoring graphene sheets to specific shapes during annealing in hydrogen atmosphere around 900°C, the high temperature, high pressure environment being crucial to the production of $SiO_x$ particles from the surface of a $Si/SiO_2$ wafer.[12] The previously un-reported drilling mechanism observed on pristine samples is therefore quite different and may depend crucially on the vacancy formation energy barrier in case of a single Si ad-atom, Si cluster, $SiO_2$ molecules and/or $SiO_2$ cluster. We found computationally that single Si ad-atoms and crystalline Si clusters have similar vacancy formation energies (8.33eV and 8.36eV, respectively). These are close to the energy required for the formation of a single vacancy in graphene by irradiation (7eV)[42] and it can therefore be concluded that the presence of Si ad-atoms or crystalline clusters should not lead to any drilling behaviour. Similarly, un-passivated quartz (ordered $SiO_2$) was revealed to be relatively 'safe' for graphene: the energy required for the migration of a single oxygen from $SiO_2$ to the graphene surface is 3.14 eV (much higher than the energy required for the oxidation of graphene reported in the literature), while the unzipping of graphene along defect 'lines' of carbon monoxide molecules[43] is unlikely as the required energies for a single and a pair of epoxy groups are 6.94 eV and 9.69 eV, respectively. On the other hand, disordered silicon oxide molecules and clusters provide a plausible solution as both have a tendency to form metastable intermediate $Si_xO_yC_z$ structures.[44] The energies required for this process are 3.13 eV and 2.46 eV for silicon oxide molecules and silicon oxide clusters, respectively, which are very much comparable to the 2.5 eV energy formation of a mono-vacancy in the presence of Ni[26] (see supplementary material for details on the modelling parameters, fig. S5). We therefore propose that as in the case of metal-deposited graphene, disordered $SiO_x$ molecules find themselves dragged to the edge of the contamination layer, where they interact with the graphene sheet. Upon formation of an initial unstable $Si_xO_yC_z$ structure, as suggested by our calculations, a hole appears and grows through the same mechanism, Si atoms or $SiO_x$ clusters being drawn energetically to the exposed edges.

Finally, we point out that the exact role of the electron beam in either the metal-mediated drilling or in the Si case could not be directly elucidated from our sole observations. However, the formation of holes at the edge of the contamination layers away from the region being imaged (up to tens of nm) and the excellent conduction properties of graphene[45] point to a heat transfer mechanism, whereby local heating as a result of the irradiation by the beam is sufficient to overcome the defect formation energy barriers.

**Conclusions**

Using atomic resolution HAADF imaging and identification of single atoms by EELS, we have observed etching of suspended, single-layer graphene upon which metal atoms had been deposited. Etching occurred with all employed metals (Cr, Ti, Pd, Ni and Al) apart from gold. It also occurred in pristine samples with (unintentional) Si contamination. Metal clusters nucleate initially exclusively on hydrocarbon contamination. Nano-scale holes form in locations where metal atom clusters sit at the border of the contamination with pristine graphene, following the drag of individual metal atoms onto the pristine graphene surface during the e-beam scan. Theoretical modelling predicts that vacancy formation energies in graphene are substantially lowered in the presence of metal atoms. Although according to such calculations the presence of oxygen is not required in this metal-mediated vacancy formation, we suggest oxygen is present and indeed assists C-C bond dissociation *via* graphene oxidation. We derive this from the fact that vacancy formation in the presence of Si-oxide has much lower predicted energies -similar to these calculated for vacancy formation in the presence of Ni- than for Si-atoms or Si-clusters alone. The presence of O is further indicated by the fact that holes form at the edges of hydrocarbon contamination, added to observations that atoms from the deposits are being dragged onto the clean graphene. Lastly, etching does not occur in the presence of gold atoms, which do not oxidise, although the predicted vacancy formation energy is lowered. The role of the scanning electron probe is not fully established, but since hole formation also occurred in regions adjacent to but not directly in the e-beam scanned area, we suggest that the e-beam acts as a heat source.

**Materials and Methods**

The monolayer CVD grown graphene membranes were transferred to the TEM support grids using a standard wet chemistry methodology. Conventional TEM was used to the assess quality of the produced films: electron diffraction data confirmed near perfect coverage of the entire TEM grids with a single layer graphene sheet.[46] The samples were then placed either into an electron beam evaporator (for depositing Au, Cr, Ti and Pd) or a thermal evaporator (for depositing Al and Ni). In all cases, the same amount of material (2 Å of Au, Cr, Ti, Pd, Ni and Al) was deposited with a precisely calibrated rate of 0.1 Å/s, in a custom-made deposition chamber whose pressure ranged from $10^{-6}$ to $10^{-8}$ torr during the evaporation. While Si atoms were not introduced deliberately, they were consistently observed as a widely present contaminant both on metal-deposited samples and on pristine graphene samples. The presence of Si (and $SiO_x$) contaminants on graphene has been widely reported for both CVD-grown and exfoliated graphene[41]. The support films, another possible origin for the Si contaminants, were all but ruled out as a source in our case: Si single atoms and $SiO_x$ clusters were indeed consistently observed on as-prepared pristine graphene samples transferred onto different types of support grids (lacey carbon film, holey carbon film and Quantifoil™) purchased from different suppliers.

All electron micrographs were acquired at the SuperSTEM Laboratory on a Nion UltraSTEM100™ dedicated scanning transmission electron microscope equipped with a cold field emission gun with a native energy spread of 0.35 eV and operated at 60 kV to prevent knock-on damage to the graphene samples. This instrument has an ultra-high-vacuum (UHV) design throughout, allowing pressures at the sample of below $5 \times 10^{-9}$ torr. The beam was set up to a convergence semi-angle of 30 mrad with an estimated beam current of 45 pA at the sample. In these operating conditions the estimated probe size is 1.1 Å. The high angle annular dark field (HAADF) detector used to record the Z-contrast images had inner and outer radii of 86 mrad and 190 mrad respectively. Detectors with lower inner angles (medium and low angle annular dark field detectors, MAADF and LAADF) can provide an increased signal-to-noise in images of low atomic number materials while retaining good signal interpretability for ultra thin samples (there is no dynamical effect for samples one atom thick) and they have been used recently to great effect for atom-by-atom chemical analysis.[47] As this study is concerned with impurity atoms deposited on top of the graphene samples, HAADF imaging was used throughout to avoid potential non-linearity effects in MAADF images, thus retaining the approximate $Z^2$ dependence of the imaging process whereby the intensity recorded with the probe positioned on an atomic site is approximately proportional to the square of the average atomic number Z of this site.[47] Electron energy loss spectra were recorded on a Gatan Enfina spectrometer with acquisition times between 1 s and 2 s (as specified in the text) for point spectra. The spectrometer acceptance semi-angle was calibrated at 33 mrad.

Interactions between graphene and silicon adatom (clusters) or $Si_xO_y$ clusters were further studied using a first-principles calculation method implemented in the SIESTA code,[48] as was done in previous work.[25, 26, 40] In modelling these interactions, the generalized gradient approximation (GGA-PBE)[49] was adopted to describe the exchange-correlation energy, which has been used in understanding graphene destruction[26] and oxidation.[25, 40] Another theoretical consideration is the interaction between graphene and a quartz substrate. In this case, the local density approximation (LDA)[50] instead of the GGA-PBE is used because the latter fails to describe the weak graphene-substrate interactions.[51] The atomic positions were fully relaxed within the maximum force of 0.04 eV/Å on individual atoms. The ion cores are described by norm-conserving non-relativistic pseudo-potentials[52] with cut off radii 1.90, 1.15 and 1.25 a.u. for Si, O and C respectively, and the wavefunctions are expanded with localized atomic orbitals (a double-ζ plus polarization basis set). All calculations were carried out with an energy mesh cut-off of 360 Ry. We used a rectangular shaped supercell containing 48 carbon atoms to model interactions between graphene and silicon atoms or clusters.[40] For the modelling of the interaction of graphene with an un-passivated quartz (ordered $SiO_2$) surface we used a graphene supercell containing 32 carbon atoms over 9 Si atomic layers of α-quartz previously used in Ref. [51]. For these two models we used the k-point mesh of 8×6×1 and 4×4×1 in the Monkhorst-Pack scheme[53] respectively.

**Acknowledgments**

This work was supported by the EPSRC (UK). We acknowledge computational support from the CAC of KIAS. Y.-W. S. was supported by the NRF grant funded by the government of Korea, MEST (Quantum Metamaterials Research Center, R11-2008-053-01002-0).

**Supporting Information available**

A document is available on-line with supporting information for the work presented in this article. It contains details about the sample screening procedure, additional images of the etching process, and a detailed description of the modelling parameters and assumptions used for the *ab initio* calculations. This material is available free of charge *via* the Internet at http://pubs.acs.org.

# Supporting Information

# Direct Experimental Evidence of Metal-Mediated Etching of Suspended Graphene

Quentin M. Ramasse[1,*], Recep Zan[2,3], Ursel Bangert[2], Danil W. Boukhvalov[4], Young-Woo Son[4] and Konstantin S. Novoselov[3]

[1]SuperSTEM Laboratory, STFC Daresbury Campus, Daresbury WA4 4AD, U.K.

[2]School of Physics and Astronomy, The University of Manchester, Manchester, M13 9PL, U.K.

[3]School of Physics, The University of Manchester, Manchester, M13 9PL, U.K.

[4]School of Computational Sciences, Korea Institute for Advanced Study, Seoul 130-722, Korea.

[*]To whom correspondence should be addressed, qmramasse@superstem.ac.uk

All TEM samples were screened in a conventional TEM (Tecnai F30) prior to metal evaporation to assess the quality of graphene coverage of the grids and to identify the number of layers by using electron diffraction.

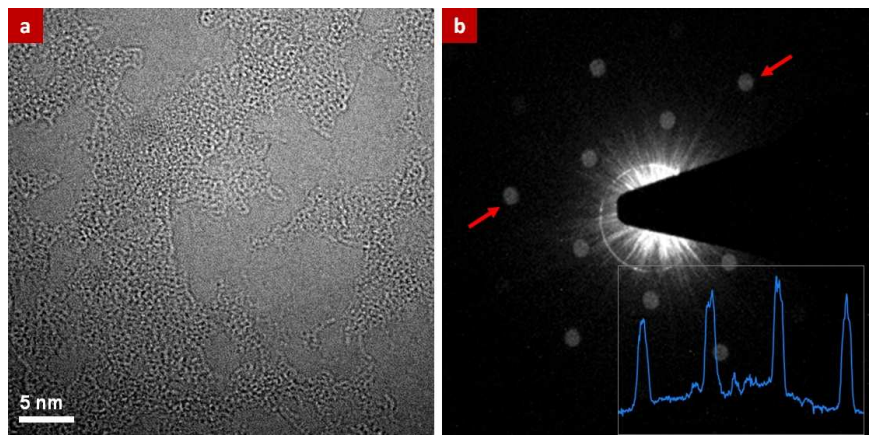

**Figure S1.** *Bright field HRTEM image (a) and diffraction pattern (b) from the imaged area of a graphene sample prior to metal deposition. The intensity profile in the inset is*

*taken between the two red arrows as indicated and confirms the presence of monolayer graphene.*

Over 90% of the surface of the samples was consistently identified as a monolayer by comparing first and second ring spot intensities in the diffraction pattern, as illustrated in figure S1.

Holes were observed to form under the beam at the edge of the contamination layer where the deposited metals (0.2nm of Ni, Pd, Ti or Al) sit preferentially. These contamination layers are believed to be due in part to carbonaceous residues from the wet chemistry methodology used to transfer the graphene layers onto TEM grids, but mostly to post-transfer exposure of the graphene to air[1]. This drilling/etching behaviour is extremely dynamic and was captured as time series with large numbers of consecutive images with very short pixel times. Many such series were acquired for all metal deposited samples: Figure S2 illustrates the process with stills from a series of images of Ni-mediated drilling. The initial hole (top) can be seen to open up all along the metal clusters with single metal atoms decorating its edges before being 'drilled' away.

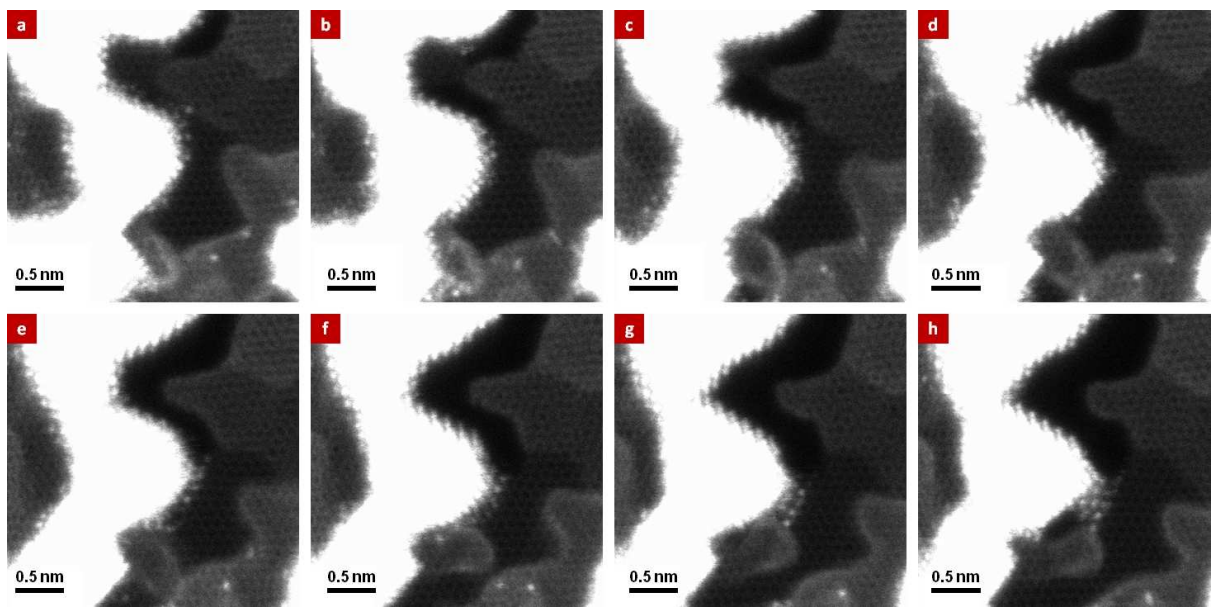

**Figure S2.** *Stills from a time series showing the metal-mediated formation of hole at the edge of a Ni cluster. The time elapsed between each image (a)-(h) is approximately 5s. To*

*improve signal-to-noise, the images shown are the sum of 10 consecutive members of the time series.*

The drilling or etching behaviour continues as long as new metal atoms jump to the edge of the holes from the neighbouring clusters, which act as a reservoir. The images in figure S3 are taken from another time series from the Pd-deposited sample. The hole is initially enlarged, with bright atoms decorating its edges jumping from the neighbouring cluster (figure S3(a)). This jumping process is too fast to be captured in a single frame, even with short pixel dwell times, so a detailed analysis of the atomic movement is difficult to provide. But bright horizontal streaks in the images, corresponding to the same atom being captured at different positions by the beam along a line, provide evidence of these jumps[1] and show that the metal atoms can travel at least a few nm within a single image line (a few ms). After a few seconds, the size of the hole stabilises (figure S3(b)) as the reservoir of mobile ad-atoms is exhausted. The dotted outlines on figure S3 (b) and (c) illustrate how the edges of the hole then simply reconfigure without any further expansion, in spite of the long time elapsed (40s between fig. S3 (b) and (c)).

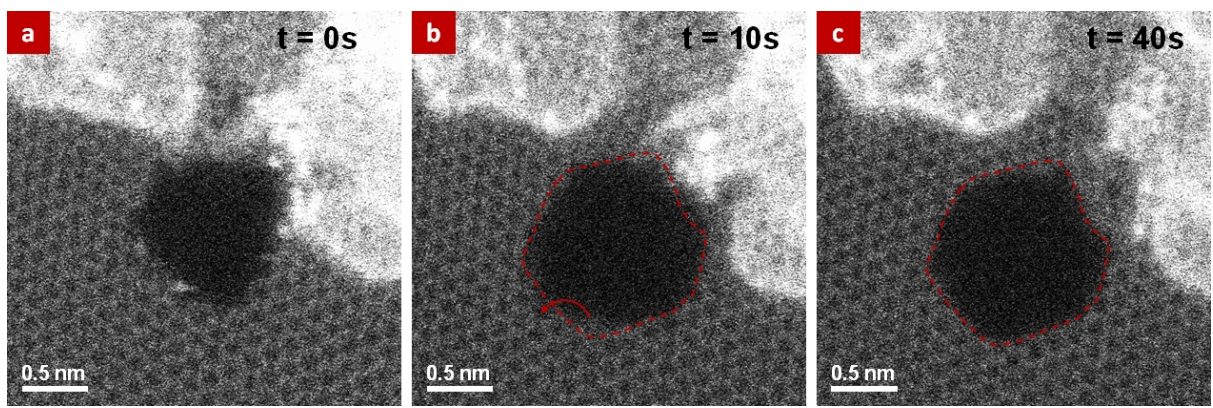

**Figure S3.** *(a) A hole formed in monolayer graphene at the edge of a Pd cluster is further enlarged as metal atoms jump onto the exposed edge, until the reservoir of impurity atoms is exhausted (b), approximately 10s later. The hole merely reshapes dynamically (c) in spite of a further 40 scans (approximately 20s), as indicated by the dotted outline of the hole and the arrow.*

The type of metal mediating the etching process did not appear to influence the shapes of the holes formed, nor the speed or ease with which their sizes grew. Such relationships may not be entirely ruled out but the variations in metal cluster sizes and distribution, and more generally the large number of parameters that would need to be controlled (exact size of the clean graphene patch being drilled, geometry of the contamination patch, proximity to the edge of the metal reservoir, etc...) make it difficult to draw any statistically significant conclusion.

This metal mediated etching was observed for all studied metals with the exception of Au, in spite of extensive observations due to the importance of Au as a potential contact in the metal-graphene system[3]. Figure S4 shows simultaneously acquired bright field and HAADF STEM images of Au-evaporated monolayer graphene, obtained in conditions identical to the other metal-deposited samples, but showing no hole formation.

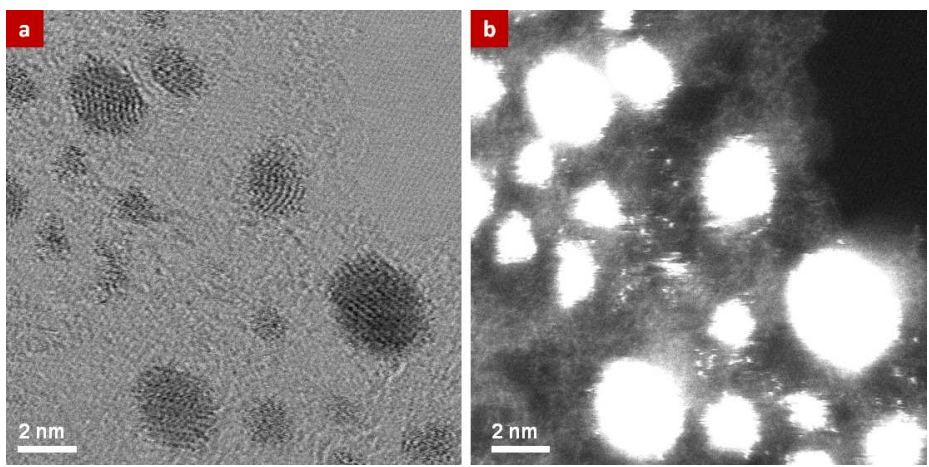

**Figure S4.** *Bright field (a) and HAADF (b) images of Au-evaporated monolayer graphene, observed in the same conditions as all other metal-deposited samples. No hole formation was observed with Au.*

Although the destruction of graphene by metal ad-atoms was predicted by theoretical calculations in the literature[4], we also observed this etching process on 'pristine' samples, consisting of monolayer graphene processed in identical conditions but on which no metal was deposited. In this case, Si atoms were found to decorate the holes and to mediate their formation. $Si/SiO_2$ are commonly found contaminants on pristine graphene[5] but were not previously thought of as potentially deleterious. Additional calculations were thus carried out to shed light on this behaviour.

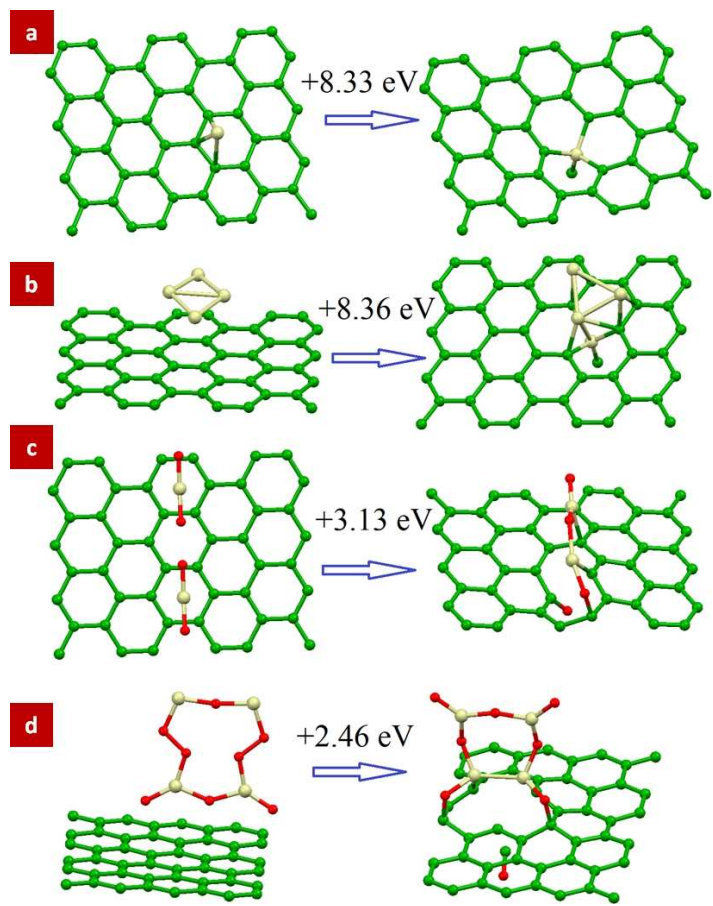

**Figure S5.** *Optimised atomic structure and energetics for initial (left column) and final (right column) steps of the formation of C vacancies in the presence of (a) a single Si atom, (b) a metallic Si cluster, (c) a $SiO_2$ molecule and (d) a $SiO_2$ cluster.*

When modelling the formation of vacancies in graphene in the presence of a single silicon ad-atom (Fig. S5(a)) or of a pure silicon cluster (fig. S5(b)), we calculated that the energies required for are in both cases higher than the energy required for the formation of a single vacancy by mere irradiation (about 7 eV). Any mechanism involving pure Si is therefore unlikely to result in the drilling behaviour we observed and it can be concluded that pure silicon is safe for graphene.

Another model involves the possible migration of oxygen atoms from unpassivated quartz (arising from the sample preparation) to the graphene surface. The energy required here for the activation of graphene with the formation of a carbon monoxide molecule is for a single epoxy group 6.94 eV, and 9.69 eV for a pair of epoxy groups[6]. Again, these calculations suggest that perfect $SiO_2$ is rather safe for graphene.

Lastly we modelled the vacancy formation energy in the presence of disordered silicon oxide, either in the form of $SiO_2$ molecules (fig. S5c) or of $SiO_2$-based loose clusters (fig. S5d). The energies required here are much lower and the interaction of graphene with the non-stoichiometric surface of $SiO_2$ nanoparticles could provide a path for the formation of metastable $Si_xO_yC_z$ structures and thus the unzipping of the graphene structure.